\documentclass[fleqn,usenatbib]{mnras}

\usepackage{amsmath}
\usepackage[amssymbols]{newtxmath}   
\usepackage{newtxtext}
\usepackage{booktabs}
\usepackage{longtable}
\usepackage{lipsum}
\usepackage{xcolor}
\usepackage{hyperref}
\usepackage{cuted}
\usepackage{tablefootnote}

\usepackage[T1]{fontenc}

\DeclareRobustCommand{\VAN}[3]{#2}
\let\VANthebibliography\thebibliography
\def\thebibliography{\DeclareRobustCommand{\VAN}[3]{##3}\VANthebibliography}

\usepackage{graphicx}	
\usepackage{amsmath}
\usepackage{stfloats}

\title[Mrk 421 NuSTAR]{Extreme Hard X-ray Variability in the TeV Blazar Mrk 421: Insights from \textit{NuSTAR} Observations}

\author[Bhatta \& Bhatta]{
Gopal Bhatta$^{1}$\thanks{E-mail: g.bhatta@ia.uz.zgora.pl}
and Eshaan Bhatta$^{2}$
\\
$^{1}$Janusz Gil Institute of Astronomy, University of Zielona Góra, ul. Szafrana 2, 65-516 Zielona Góra, Poland\\
$^{2}$Independent researcher, ul. Juliusza Lea 10B, Krakow,Poland}

\date{Accepted XXX. Received YYY; in original form ZZZ}

\pubyear{\the\year{}}

\begin{document}
\label{firstpage}
\pagerange{\pageref{firstpage}--\pageref{lastpage}}
\maketitle

\begin{abstract}
In this work, we investigate the hard X-ray variability of the TeV blazar Mrk 421 over both long-term and intra-day timescales to gain insights into the particle acceleration and cooling mechanisms driving its emission. Using 27 \textit{NuSTAR} observations spanning 2012 to 2024, we conduct detailed spectral and timing analyses. Our findings reveal pronounced flux variability, with changes by up to a factor of 200 over the decade and a 10-fold increase within a period of 15 days during a flaring episode. The flux distribution exhibits a lognormal profile, and a linear root-mean-square (RMS)--flux relation suggests a multiplicative variability process. Spectral analysis indicates that a log-parabolic model best describes most observations, with a strong linear anti-correlation between the photon index and curvature parameter. We also identify a positive correlation between the synchrotron peak energy and flux, accompanied by a subtle ``harder-when-brighter'' trend. These results align with standard blazar emission models, supporting particle acceleration via shock fronts or magnetic reconnection in the relativistic jet.
\end{abstract}




\section{Introduction}
\label{introduction}

Blazars constitute the most energetic class of radio-loud active galactic nuclei (AGN) in the universe. These remarkable sources are characterized by relativistic jets aligned closely with the observer’s line of sight, producing intense non-thermal emission across the entire electromagnetic spectrum, from radio to high-energy $\gamma$-rays. Relativistic jets in AGN, powered by supermassive black holes, are aligned close to our line of sight, causing Doppler boosting that amplifies observed emission and induces rapid variability across all wavebands \citep{Meier12, Urry1995}. These jets, accelerating particles to near-light speeds, make blazars ideal for studying extreme astrophysical processes and high-energy particle acceleration \citep{Blandford2019}.

The broadband spectral energy distribution (SED) of blazars typically exhibits a characteristic double-peaked structure in the frequency-flux plane. The lower-frequency peak, spanning radio (mm-wave) to X-ray bands, arises from synchrotron emission generated by relativistic particles accelerated within the jet’s magnetic field. The origin of the higher-frequency peak, extending from X-rays to TeV energies, remains a topic of active debate. Leptonic models attribute this emission to inverse-Compton scattering of soft seed photons by high-energy electrons \citep{Maraschi1992, Bloom1996}. In the synchrotron self-Compton (SSC) scenario, the same electron population responsible for synchrotron emission upscatters these photons to higher energies. Alternatively, the external Compton (EC) scenario proposes that seed photons originate from external AGN components, such as the accretion disk \citep{Dermer1993}, broad-line region \citep{Sikora1994}, or dusty torus \citep{Blazejowski2000}. In contrast, hadronic models suggest that high-energy emission results from interactions between relativistic protons and ambient radiation fields \citep{Mannheim1992, Aharonian2000, Mucke2003}.

Blazars are classified into two main categories: flat-spectrum radio quasars (FSRQs) and BL Lacertae (BL Lac) objects. FSRQs are generally more luminous, displaying prominent optical emission lines and synchrotron peaks at lower frequencies. BL Lac objects, by contrast, are less luminous, with weak or absent emission lines and synchrotron peaks at higher frequencies. BL Lacs represent an extreme subclass, with synchrotron emission extending to soft X-rays (up to $10^{16}$ Hz) and inverse-Compton emission spanning hard X-rays to TeV energies. These sources typically lack strong circumstellar photon fields and are thought to accrete matter at relatively low rates \citep[see][and references therein]{2018A&A...619A..93B}. BL Lac objects are further subdivided based on their synchrotron peak frequency ($\nu_{\text{sy}}$): low-synchrotron-peaked (LSP, $\nu_{\text{sy}} < 10^{14}$ Hz), intermediate-synchrotron-peaked (ISP, $10^{14}$ Hz $< \nu_{\text{sy}} < 10^{15}$ Hz), and high-synchrotron-peaked (HSP, $\nu_{\text{sy}} > 10^{15}$ Hz) sources \citep{2010ApJ...722..520A}.

Blazars exhibit remarkable variability across all frequencies, from radio to TeV wavebands, spanning timescales from minutes to decades. Multi-wavelength (MWL) variability has emerged as one of the defining properties of blazars, with variability studies playing a significant role in advancing our understanding of these sources \citep[e. g., see][and references therein]{Navaneeth2025}. This variability can be characterized as a combination of three primary components: a general red noise-like behavior dominating the broadband power spectral density, occasional high-amplitude flares, and, in certain blazars, quasi-periodic oscillations with diverse characteristic timescales \cite[e. g., see][and the reference therein]{2023MNRAS.520.2633B,2019MNRAS.487.3990B,2017ApJ...847....7B,Gupta2018,2016ApJ...832...47B}.

Markarian 421 (Mrk 421), a high-synchrotron-peaked BL Lac object at a redshift of $z = 0.031$, is one of the closest and brightest TeV-emitting blazars, making it an ideal target for studying relativistic jet physics in AGN. Its non-thermal emission spans the entire electromagnetic spectrum, from radio to very-high-energy (VHE) $\gamma$-rays, exhibiting significant flux and spectral variability. Consequently, Mrk 421 has been the focus of extensive multi-wavelength (MWL) observational campaigns, covering radio to TeV energies, which have revealed complex variability patterns throughout its observational history \citep{2025A&A...695A.217M,2025A&A...694A.195A, 2022MNRAS.513.1662M, 2016ApJ...819..156B}. These broadband efforts have produced invaluable datasets, providing crucial constraints for theoretical models of jet emission mechanisms \citep{2011ApJ...736..131A, 2005ApJ...630..130B, 2008ApJ...677..906F}. Studies utilizing (quasi-)simultaneous X-ray and $\gamma$-ray observations have frequently reported correlated variability between these bands \citep{ 1999ApJ...526L..81M, 2003A&A...402..929B, 2005A&A...443..397B, 2009A&A...501..879T, 2010PASJ...62L..55I, 2014A&A...570A..77P, 2016ApJ...827...55K}. In the X-ray regime, Mrk 421’s variability has been extensively studied using instruments such as XMM-Newton, Swift, NuSTAR, NICER and AstroSAT. These analyses reveal strong correlations between soft and hard X-ray emission, with spectra often described by power-law or curved models \citep{Kizhakkekalam2026,2025JHEAp..45..438A, 2016ApJ...831..102K, Pandey2017, 2022ApJS..262....4N, 2001A&A...365L.162B}. Observations with Chandra have also provided evidence for warm-hot intergalactic medium (WHIM) filaments along the line of sight, confirming their role as reservoirs of baryonic matter in the cosmic web \citep{2005ApJ...629..700N, 2006ApJ...652..189K}. Detailed studies of flaring episodes have revealed interesting spectral and temporal features, including spectral hysteresis loops in the flux–hardness ratio plane and evidence of both hard and soft lags \citep{2004A&A...424..841R, 2009A&A...501..879T, Khatoon2022}.

Hard X-ray emission from blazars--and AGNs in general-- detected by several X-ray telescopes, such as Swift/BAT \citep{2018ApJS..235....4O} and INTEGRAL \citep{2022MNRAS.510.4796K}. In particular, the hard X-ray variability properties of blazars observed by NuSTAR have been discussed in detail \citep{2022MNRAS.514.3179M,2018A&A...619A..93B}. The presence of hard X-ray excess in blazars is especially significant in light of the potential association between AGNs and high-energy neutrinos. Such an association can be naturally explained by hadronic models, in which secondary emission from hadronic processes—particularly proton–photon ($p\gamma$) and Bethe–Heitler (BH) interactions \citep[e.g., see][]{2012JPhCS.355a2011R}—can significantly contribute to, or even dominate, the X-ray emission \citep[see][]{2019MNRAS.483L..12C,2013MNRAS.434.2684M}. This would suggest that the observed hard X-ray excess may be indicative of underlying hadronic emission mechanisms. In such scenario, synchrotron emission from secondary pairs—produced via $p\gamma$ and BH interactions—can account for the hard X-ray excess. These high-energy photons may also contribute to the very high energy (VHE) band through inverse Compton scattering, potentially leading to a correlation between X-ray and VHE variability. As a result, hadronic emission would naturally show a lack of correlation with emission dominated by leptonic processes.
In addition to these spectral and variability signatures, recent IXPE observations of Mrk 421 have revealed strong and variable X-ray polarization consistent with energy-stratified, multizone synchrotron emission \citep{DiGesu2022,DiGesu2023,Kim2024,Maksym2025}, providing a complementary diagnostic for constraining leptonic and hadronic high-energy emission scenarios \citep{Zhang2024}.

In this paper, we present a comprehensive analysis of all available \textit{NuSTAR} observations of Mrk 421, focusing on timing and spectral analyses to explore the variability properties of this blazar in the hard X-ray regime. Our study aims to explore the physical processes driving blazar emission by exploring the relationships among variability, mean flux, and spectral properties, revealing insights into the innermost regions of blazar jets, which remain mostly inaccessible to direct observation.

We organize the paper as follows: In Section 2, we describe the 27 \textit{NuSTAR} observations and the data processing methods. In Section 3, we present the methods used for timing and spectral analyses based on light curves and spectra, and we discuss the results. In Section 4, we discuss key observational findings, including rapid flux and spectral variability, and a correlation between higher flux and harder spectra interpreting these in the context of current blazar models. Finally, in Section 5, we summarize our conclusions.

\begin{figure*}
\centering
\includegraphics[width=\textwidth]
{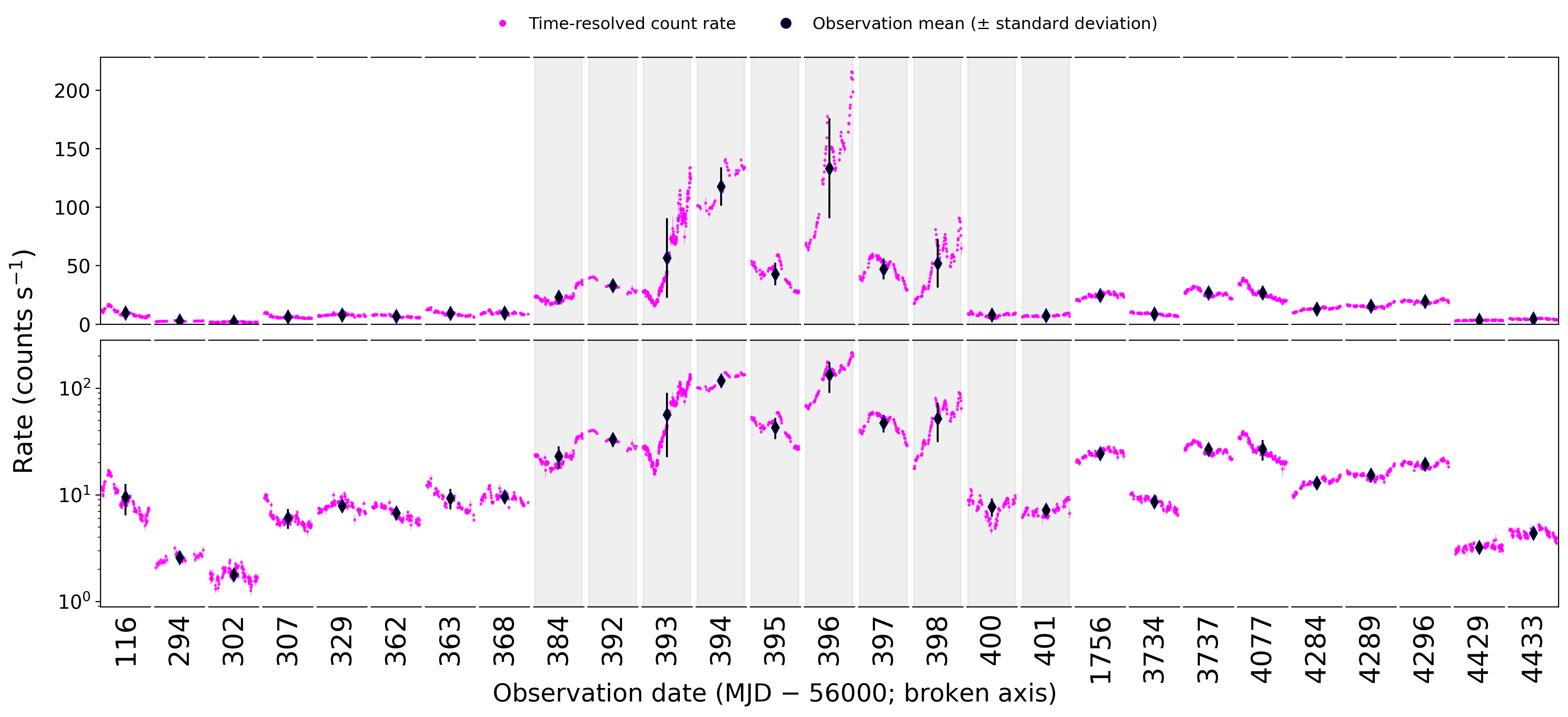}
\caption{The upper and lower panels show the long-term NuSTAR light curve on linear and logarithmic count-rate scales, respectively. Magenta points represent the time-resolved count rates, while the larger black symbols and error bars show the mean and standard deviation for each observation. Observation dates are given as MJD \(-\) 56000, with the corresponding calendar dates listed in Table \ref{tab1}. The logarithmic representation highlights the low-flux states.  \label{fig1}}
\end{figure*}

\begin{table*}
\caption{Properties of \textit{NuSTAR} observations of the blazar Mrk 421. Columns 2, 3, 4, and 5 list the observation date, MJD minus 56000 (at 00:00 UTC on the listed date), observation ID, and exposure time (in seconds), respectively, while Columns 6 and 7 provide the median flux (in counts/s)  and the variability parameter with its uncertainty (in percent), respectively. \label{tab1}}
\centering
\begin{tabular}{llllllc} 
\toprule
Serial & Date & MJD$-56000$ & Obs. ID & Exposure(s) & Median (c/s) & $P_{\rm var} \pm \Delta P_{\rm var}$ (\%) \\
(1) & (2) & (3) & (4) & (5) & (6) & (7)\\
\midrule
1    & 2012-07-08 & 116  & 10002016001 & 24884      & 8.72                & 30.48 $\pm$ 1.05 \\
2    & 2013-01-02 & 294  & 60002023002 & 9152       & 2.58                & 8.85 $\pm$ 0.55 \\
3    & 2013-01-10 & 302  & 60002023004 & 22631      & 1.73                & 14.08 $\pm$ 1.09 \\
4    & 2013-01-15 & 307  & 60002023006 & 24181      & 5.64                & 18.10 $\pm$ 0.77 \\
5    & 2013-02-06 & 329  & 60002023010 & 19302      & 7.79                & 9.81 $\pm$ 0.36 \\
6    & 2013-02-12 & 335  & 60002023012 & 14775      & 9.14                & 20.49 $\pm$ 0.73 \\
7    & 2013-03-11 & 362  & 60002023018 & 17472      & 6.80                & 14.12 $\pm$ 0.57 \\
8    & 2013-03-17 & 368  & 60002023020 & 16554      & 9.48                & 9.45 $\pm$ 0.32 \\
9    & 2013-04-02 & 384  & 60002023022 & 24767      & 21.38               & 21.30 $\pm$ 0.49 \\
10   & 2013-04-10 & 392  & 60002023024 & 5757       & 31.82               & 15.26 $\pm$ 0.30 \\
11   & 2013-04-11 & 393  & 60002023025 & 57507      & 42.88               & 73.38 $\pm$ 1.32 \\
12   & 2013-04-12 & 394  & 60002023027 & 7629       & 113.73              & 13.93 $\pm$ 0.15 \\
13   & 2013-04-13 & 395  & 60002023029 & 16508      & 44.72               & 24.79 $\pm$ 0.40 \\
14   & 2013-04-14 & 396  & 60002023031 & 15605      & 145.84              & 38.04 $\pm$ 0.34 \\
15   & 2013-04-15 & 397  & 60002023033 & 17276      & 49.26               & 20.23 $\pm$ 0.30 \\
16   & 2013-04-16 & 398  & 60002023035 & 20278      & 57.20               & 49.58 $\pm$ 0.70 \\
17   & 2013-04-18 & 400  & 60002023037 & 17795      & 8.09                & 20.58 $\pm$ 0.75 \\
18   & 2013-04-19 & 401  & 60002023039 & 15958      & 6.99                & 12.39 $\pm$ 0.48 \\
19   & 2017-01-03 & 1756  & 60202048002 & 23691      & 24.68               & 8.01 $\pm$ 0.17 \\
20   & 2022-06-04 & 3734  & 60702061002 & 21277      & 8.75                & 12.27 $\pm$ 0.42 \\
21   & 2022-06-07 & 3737  & 60702061004 & 23280      & 26.21               & 10.28 $\pm$ 0.21 \\
22   & 2023-05-13 & 4077  & 80801643002 & 44571      & 25.57               & 20.34 $\pm$ 0.37 \\
23   & 2023-12-06 & 4284  & 60902024002 & 21293      & 13.21               & 11.65 $\pm$ 0.26 \\
24   & 2023-12-11 & 4289  & 60902024004 & 21235      & 15.22               & 7.57 $\pm$ 0.16 \\
25   & 2023-12-18 & 4296  & 60902024006 & 19990      & 19.33               & 5.84 $\pm$ 0.11 \\
26   & 2024-04-29 & 4429  & 60902024010 & 20967      & 3.20                & 5.12 $\pm$ 0.24 \\
27   & 2024-05-03 & 4433  & 60902024012 & 21025      & 4.33                & 7.95 $\pm$ 0.31 \\
\bottomrule
\end{tabular}
\label{tab:data_serial_fvar}
\end{table*}

\section{NuSTAR observations and data processing}

The Nuclear Spectroscopic Telescope Array (\textit{NuSTAR}) is a sensitive hard X-ray (3--79 keV) observatory equipped with two focal plane modules, FPMA and FPMB, designed to provide high-resolution imaging and spectroscopy. 

The \textit{NuSTAR} instrument achieves a spectral resolution of $\sim$400 eV FWHM at 6 keV, increasing to $\sim$1 keV at 68 keV across its 3--79 keV bandpass. Each telescope provides a field of view of $\sim$13'  and a half-power diameter of $\sim$58"  for a point source image \citep{Harrison2013}. This capability makes \textit{NuSTAR} an ideal tool for studying the hard X-ray variability and spectral properties of blazars like Mrk 421, which often exhibit complex emission in this energy range.

To characterize the hard X-ray emission from the blazar Mrk 421, we conducted a comprehensive analysis of both flux variability and spectral properties using 27 \textit{NuSTAR} observations of the source, selected based on two primary requirements: an exposure time greater than 1000 seconds and an issue flag of zero. The observation date, the corresponding Modified Julian Date (MJD)-56000, observation ID, and exposure time for each selected dataset are presented in the 2nd, 3rd, 4th and 5th columns of Table \ref{tab1}, respectively, providing a clear overview of the observational parameters used in this study.

Archival hard X-ray observations of the blazar source were retrieved from the NuSTAR Master Catalog\footnote{\url{https://heasarc.gsfc.nasa.gov/W3Browse/nustar/numaster.html}} for analysis.
Data reduction was performed using the \textit{NuSTAR} Data Analysis Software (NuSTARDAS) package version 1.3.1, integrated with HEASOFT\footnote{\url{https://heasarc.nasa.gov/lheasoft/}} version 6.33 and the calibration database CALDB version 20220331. All observations were processed with the standard \textsl{nupipeline} script to produce calibrated and cleaned event files. Source emission was extracted from a circular region of 30" radius centered on the source position, while background emission was sampled from a 60" radius circular region in a source-free area of the detector, ensuring minimal contamination while remaining on the same detector chip. Light curves were generated with a uniform time binning of 300 seconds, balancing temporal resolution with a sufficient signal-to-noise ratio for variability analysis. Spectra were extracted and re-binned using the \emph{grppha} task to ensure a minimum of 20 counts per energy channel, enabling the application of 
$\chi^2$
  statistics in spectral fitting. This rigorous data processing approach ensures reliable characterization of the source’s hard X-ray behavior, facilitating detailed comparisons with theoretical emission models.

\section{Analysis and Results}

Our investigation can be divided into two main components: timing analysis and spectral analysis, which are presented in the subsequent sections. The timing analysis explores the temporal behavior of the source across multiple timescales, while the spectral analysis examines the underlying emission mechanisms and their evolution. Together, these complementary approaches provide insights into the physical processes driving the hard X-ray emission from this well-studied blazar.\\
\subsection{Timing Analysis}
To carry out timing analysis, we processed the data to generate intra-day light curves, as detailed in Section 2, and present them in Figure 1 in chronological order, corresponding to the observations listed in Table 1. The upper panel of Figure 1, displayed on a linear y-scale, reveals a dramatic flux change over the entire period, while the lower panel, using a logarithmic y-scale, emphasizes the intra-day variability, particularly in the low flux states. Building on these light curves, we systematically analyzed the flux variability properties across both intra-day and long-term timescales. The intra-day analysis focused on characterizing rapid fluctuations within individual observing epochs, while the long-term investigation examined variability patterns spanning the entire observational campaign. This dual-timescale approach allows us to probe both the rapid, local processes occurring within the jet and the more sustained evolutionary trends in the emission properties of Mrk 421.
\subsubsection{Flux variability}
As the flux state of the source varied across the observation epochs, we computed the median flux for each of the 27 intra-day hard X-ray light curves, as presented in the sixth column of Table \ref{tab:data_serial_fvar}. Across the observations, the median flux exhibits a wide range, from a low of 1.73 counts/s on 2013-01-10 to a high of 145.84 counts/s on 2013-04-14, with an average of 26.72 counts/s over the dataset. This broad span highlights the source’s dynamic behavior in the hard X-ray band, capturing both its baseline and peak activity levels. Notably, the maximum median flux of 145.84  counts/s on 2013-04-14 indicates a significant flaring event, one of the most intense recorded for this source in this energy regime \citep{2015ApJ...811..143P}. The observed flux variations provide valuable insights into the source’s energetic processes over time.

While a light curve's variance provides an initial measure of variability, it includes both intrinsic source fluctuations and measurement uncertainties. To isolate the source's true physical variability, we must subtract the variance contribution from measurement uncertainties in the count rates. This corrected metric, known as excess variance, effectively quantifies the intrinsic variability by eliminating statistical noise from the measurement process.
For a light curve comprising $N$ measured flux values $X_{i}$, each with associated finite uncertainties $\sigma_{\rm {err,i}}$  due to measurement errors, and given $S^{2}$ as the sample variance of the light curve, the excess variance is computed using the following relation:
\begin{equation}
\label{excess}
   \sigma^{2}_{XS} =  S^{2} - \Bar{\sigma}^{2}_{\rm err},
\end{equation}
\noindent where $\Bar{\sigma}^{2}_{\rm err}$ represents square of the mean measurement errors, providing a robust measure of the source's true variability \citep{2002ApJ...568..610E,1997ApJ...476...70N}.    Fractional variability, which is related to the normalized excess variance, is calculated using the relation,

\begin{equation}
    F_{\rm var} = \sqrt{\frac{S^{2} - \bar{\sigma}^{2}_{\rm err}}{\bar{X}^2}}
\end{equation}
\citep[see][]{1990ApJ...359...86E,2003MNRAS.345.1271V}.
The associated error in the fractional variability is obtained using,
\begin{equation}
   \sigma_{F_{\mathrm{var}}} =  \sqrt{\left(\frac{1}{\sqrt{2N}}\frac{\bar{\sigma}^{2}_{err}}{F_{var}}\frac{1}{\bar{X}^2}\right)^2 + \left(\sqrt{\frac{\bar{\sigma}^{2}_{err}}{N}}\frac{1}{\bar{X}^2}\right)^2}
\end{equation}
\citep{2015A&A...576A.126A,Bhatta2018}. 

However, for observations with non-Gaussian distributions, the mean and variance are inadequate for characterizing the sample, and the median serves as a more reliable measure than the mean. The variability observed in the X-ray light cuves are quantifies using  variability parameter defined as:

\begin{equation}
P_{\rm var} = \frac{1}{\sqrt{N}} \frac{1}{x_{\rm med}} \sqrt{\sum_i (x_i - x_{\rm med})^2 \cdot \frac{\delta_{\rm med}^2}{\delta_i^2}},
\end{equation}
where \( P_{\rm var} \) quantifies the degree to which a source’s flux deviates from its median value. The terms \( x_{\rm med} \) and \( \delta_{\rm med} \) denote the median flux and median uncertainty of the sample, respectively, while \( x_i \) and \( \delta_i \) represent individual flux measurements and their uncertainties. The weighting factor \( \frac{\delta_{\rm med}^2}{\delta_i^2} \) reduces the contribution of measurements with larger uncertainties relative to those with smaller ones. The uncertainty in \( P_{\rm var} \) is estimated as \( \Delta P_{\rm var} = \frac{\delta_{\rm med}}{x_{\rm med}} P_{\rm var} \) \citep[see][]{2024ApJ...969...98G}.

 We computed the variability parameter for all 27 intra-day hard X-ray light curves using the relations described above, with results listed in the 7th column of Table \ref{tab:data_serial_fvar}. The analysis reveals a remarkable range of variability across observations, spanning from a minimal value of approximately 5\% (observed on 2024-04-29) to an extraordinary peak of 73\% (recorded on 2013-04-11). On average, the intra-day observations exhibit a P$_ {\rm var}$ value of approximately 21\% across the entire dataset. These values highlight the dramatic variability characteristics of this source in the hard X-ray band, demonstrating both extreme fluctuations during its most active states and the wide dynamic range of variability the source experiences over time. The exceptional P$_ {\rm var}$ value of 73\% during the 2013-04-11 observation represents one of the most violent variability episodes reported for this class of object in the hard X-ray regime. Blazars, including Mrk 421, are well-known for exhibiting rapid flux variations on intra-day timescales across a broad range of electromagnetic wavebands \citep[e.g.,][]{1989Natur.337..627M, 2018A&A...619A..93B, 2022ApJS..262....4N, 2023MNRAS.518.1459P, 2023ApJ...955..121D, 2025ApJ...981..118B}. These short-timescale fluctuations, often detected in the hard X-ray regime, provide a powerful probe of the physical processes within the compact emission regions of blazar jets. Using light-travel-time arguments, the intra-day variability suggests that the emission originates from highly compact regions, likely located close to the central engine. By analyzing the variability in the hard X-ray emission of Mrk 421, we can constrain some of the important properties of these regions, including their size and relativistic kinematics, offering valuable insights into the dynamics and structure of the jet.

\begin{figure}
\centering
\includegraphics[width=9cm]{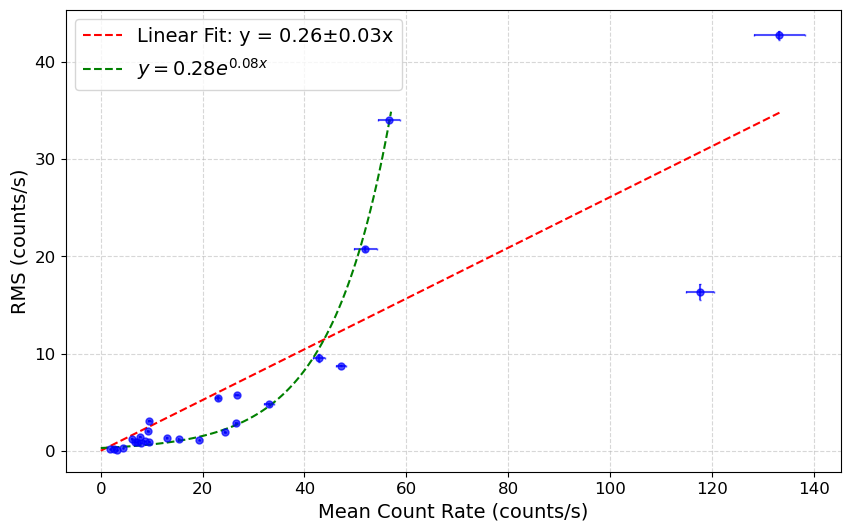}
\caption{RMS values are plotted against mean count rates of all 27 NuSTAR observations of blazar Mrk 421. The red dashed line represents the best-fit linear relation. To characterize the average trend over the low-to-intermediate count-rate range, we also fit an empirical exponential curve, shown by the green dashed line, to the 25 densely sampled observations, excluding the two observations with the highest count rates.} 
\label{rms_flux}
\end{figure}

\subsection{RMS--Flux correlation}
The root-mean-square (RMS)--flux relation in AGN reveals a linear correlation between the RMS variability amplitude and the mean flux, observed across a wide range of timescales \citep{2005MNRAS.359..345U,2004A&A...414.1091G,2012MNRAS.422.2620H}. Initially identified in the X-ray emission from AGN and X-ray binaries, this relation indicates that short-term flux variations scale multiplicatively with longer-term brightness changes, a characteristic of accretion-driven processes. The remarkable linearity persists across diverse timescales and energy bands, suggesting that variability arises from propagating fluctuations in the accretion flow, where turbulent variations in the mass accretion rate at different radii are coupled \citep{2005MNRAS.359..345U}.
\begin{figure}
\centering
\includegraphics[width=10cm]{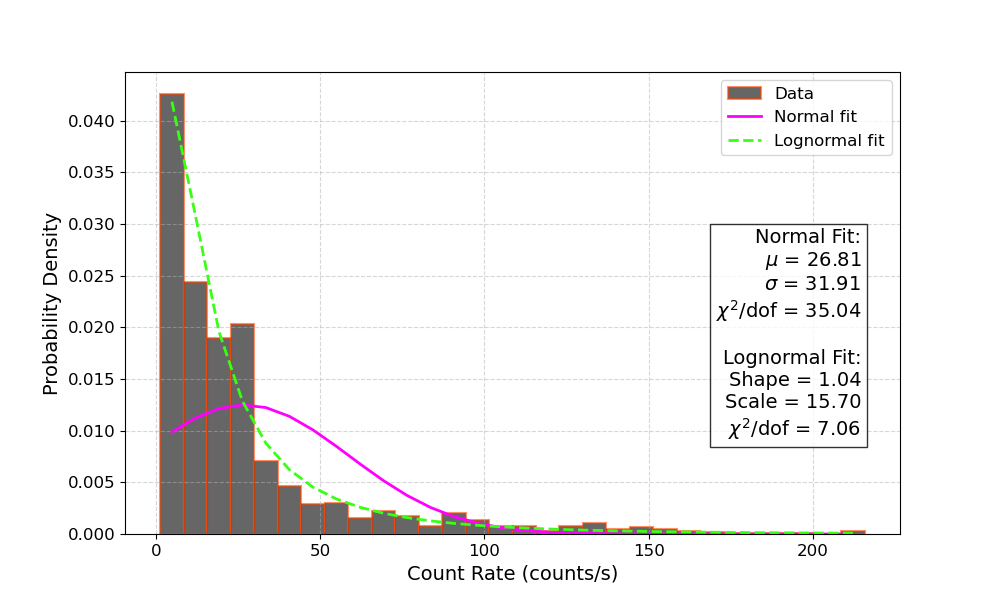}
\caption{Histogram of the count rate distribution for all 27 \textit{NuSTAR} light curves of Mrk 421, fitted with normal (magenta) and lognormal (green) probability density functions (PDFs). The inset text box provides the $\chi^2$ values and degrees of freedom from the fit statistics for each distribution. \label{fig3}}
\end{figure}

Recently, the rms–flux relation has been investigated in several blazars across different observational wavebands, with the aim of probing the disk–jet connection in jetted AGNs. The relation has been found to hold in the gamma-ray band (e.g., \cite{2023RAA....23k5011W,2020ApJ...891..120B,2017ApJ...849..138K}), X-rays \cite{Bhattacharyya2020}, and optical wavelengths \cite{Bhatta2021}. However, in long-term variability studies of the blazars 3C 273 and PKS 2155-304 based on XMM-Newton X-ray observations, the relation was not observed \citep{2023ApJ...955..121D,2025ApJ...981..118B}.

Motivated by these findings, we investigated whether the variability amplitude in the hard X-ray emission from Mrk 421 correlates with its mean flux state, aiming to further understand the nature of its variability and the physical processes driving the source's behavior. We quantified the variability using the root mean square (RMS) amplitude, defined as $\rm RMS =\sqrt{\rm excess \ variance}$, and calculated the Spearman correlation coefficient between the RMS and the mean count rate across all 27 \textit{NuSTAR} light curves. The analysis yielded a correlation coefficient of 0.77 with a p-value of 4.38x10$^{-6}$, indicating a significant positive correlation between the two quantities.
In addition, we observed that the RMS increases slowly at low count rates and more rapidly above approximately $20~\mathrm{counts~s}^{-1}$. We therefore used an exponential function as an empirical description of this nonlinear trend. The fit was restricted to the 25 observations within the densely sampled low-to-intermediate count-rate range below approximately $60~\mathrm{counts~s}^{-1}$, for which the correlation coefficient increases to 0.85, with a $p$-value of $5.42\times10^{-7}$. The two brightest observations, both above $115~\mathrm{counts~s}^{-1}$, were retained in the figure but excluded from this fit because they are separated from the main distribution by a large sampling gap and could disproportionately influence the fitted trend. Thus, the green curve is intended only as an empirical description of the average statistical behavior within the low-to-intermediate flux regime. Although the departure from a single linear trend may indicate state-dependent variability or contributions from multiple emitting regions, additional observations in the highest flux states would be required to test such interpretations.

 On the other hand, a linear RMS–flux trend—indicating that higher flux states are associated with greater variability amplitudes—is commonly observed in blazars and is often attributed to non-linear, multiplicative processes in the jet \citep{2005MNRAS.359..345U}. The relationship is visually represented in Figure \ref{rms_flux}, where a plot of the mean count rate against the RMS clearly illustrates the linear trend, reinforcing the statistical correlation.  The relation provides evidence for a connection between the source’s brightness and its variability, potentially reflecting dynamic changes in particle acceleration or energy injection within the jet of Mrk 421.

\subsubsection{Flux distribution and PDF}
Understanding the statistical distribution of X-ray flux from blazars like Mrk 421 provides critical insights into the physical processes governing their emission. In particular, the distinction between normal (Gaussian) and lognormal probability density functions (PDFs) serves as a key discriminator between theoretical models of particle acceleration and emission mechanisms. Normal distributions typically arise from additive processes, while lognormal distributions are associated with multiplicative processes, such as turbulent or cascading energy injection in the jet \citep{2005MNRAS.359..345U}. This analysis is particularly valuable for constraining the underlying physics of blazar variability.

The normal distribution is expressed as:
\begin{equation}
N(x) = \frac{1}{\sqrt{2\pi\sigma^2}} \exp\left(-\frac{(x-\mu)^2}{2\sigma^2}\right),
\end{equation}
where \(N(x)\) is the probability density function, \(x\) represents the flux value, \(\mu\) is the mean, and \(\sigma\) is the standard deviation, quantifying the distribution's spread.

The lognormal distribution is given by:
\begin{equation}
Ln(x) = \frac{1}{\sqrt{2\pi}sx} \exp\left(-\frac{(\ln x - m)^2}{2s^2}\right),
\end{equation}
where \(x\) is the positive flux value, \(m\) is the mean of the natural logarithm of \(x\), and \(s\) is the standard deviation of the logarithm. The factor \(1/x\) ensures proper normalization on the logarithmic scale. Unlike the symmetric normal distribution, the lognormal distribution is skewed, with a longer tail toward higher values, a feature often observed in blazar flux distributions.

\begin{figure*}
  \centering
  \includegraphics[width=8.1cm]
  {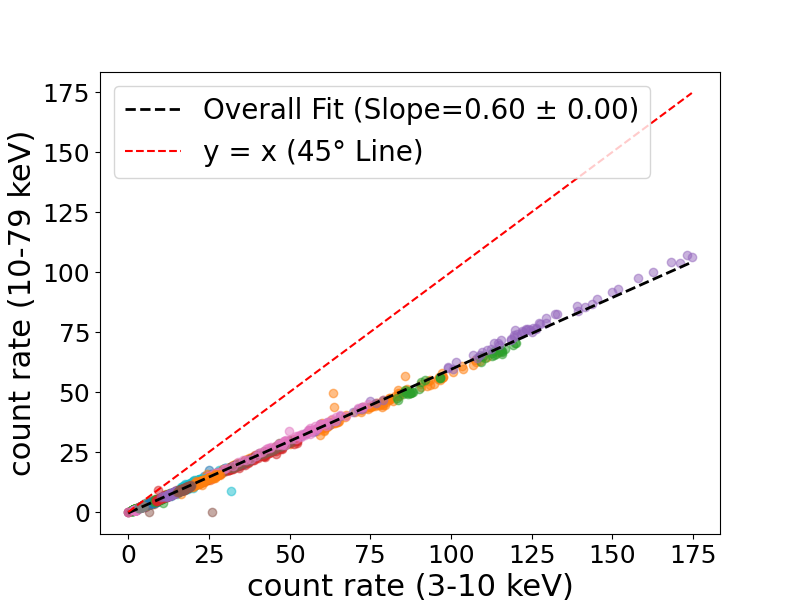}
  \includegraphics[width=7.5cm]{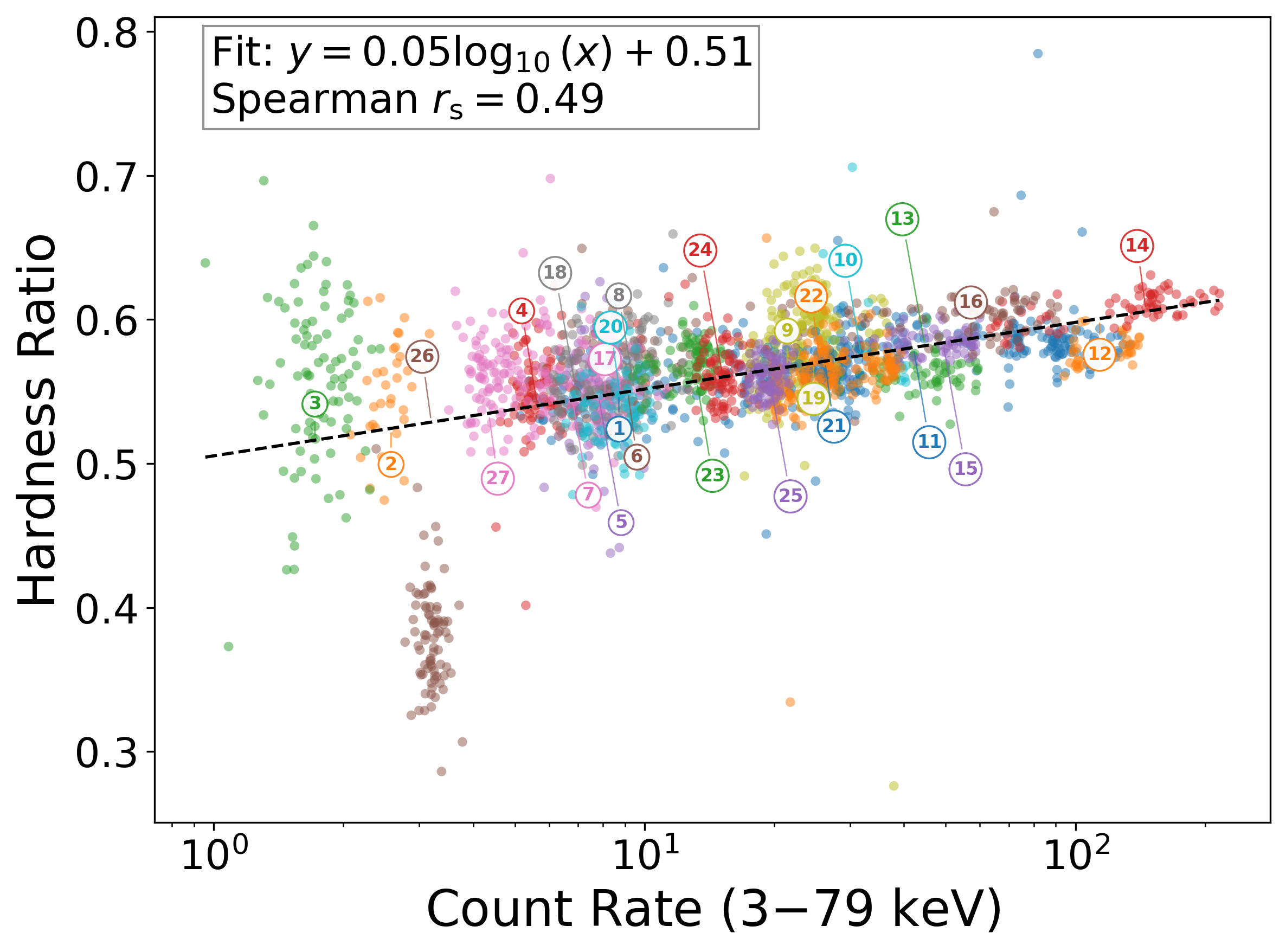}
  \caption{  \textit{Left panel}: Correlation between the \textit{NuSTAR} low-energy (3--10~keV) and high-energy (10--79~keV) count rates of Mrk 421, with the dashed red line indicating the $y = x$ relation.
  \textit{Right panel}: Distribution of the hardness ratio ( defined by Equation \ref{eq6})  as a function of count rate across the entire observation period, with distinct colors representing different observational epochs. The numbered labels, shown in the corresponding colors near the best-fit line, identify the observations in temporal order, consistent with the ordering in Table \ref{tab1}. \label{fig4}}
\end{figure*}

 We combined the all 27 intra-day light curves to construct a histogram of count rate distribution, reflecting the source’s variability across multiple observational epochs. Both normal and lognormal PDFs were fitted to the histogram using a least-squares method, with goodness-of-fit evaluated via the reduced \(\chi^2\) statistic.

Figure \ref{fig3} illustrates the histogram of flux variations, with the normal distribution (magenta line) and lognormal distribution (green line) overlaid. The lognormal PDF provides a significantly better fit, yielding a reduced \(\chi^2\) of \(\sim 7\), compared to \(\sim 35\) for the normal distribution. An inset text box in the figure reports the \(\chi^2\) values and degrees of freedom for both fits, allowing a direct comparison. This result indicates that the hard X-ray flux distribution of Mrk 421 is more consistent with lognormal behavior, suggesting multiplicative processes dominate the variability.

The preference for a lognormal distribution has been reported in previous studies using \textit{Swift}/XRT observations of Mrk 421 \cite{2020ApJS..247...27K}, and similar lognormal behavior has also been observed in \textit{XMM-Newton} observations of several other blazars \citep{2022Univ....8..578W}. In contrast, a study based on archival \textit{XMM-Newton} data spanning a decade found no clear evidence of a distinct probability distribution function (PDF) in the flux distributions of the FSRQ 3C 273 and the BL Lac object PKS 2155--304 \citep{2023ApJ...955..121D,2025ApJ...981..118B}. Complementing the X-ray results, flux distributions of several blazars—including Mrk 421—have also been shown to be consistent with a lognormal PDF in the gamma-ray \citep{2023RAA....23k5011W,2020ApJ...891..120B,2017ApJ...849..138K} and optical bands \citep{Bhatta2021}.

In blazars, the presence of a lognormal PDF in their flux distributions is often attributed to non-linear processes, potentially driven by turbulent energy injection or cascading mechanisms within the relativistic jet \citep{Shah2018}. This contrasts with additive processes, often associated with shot-noise-like behavior and characterized by a normal distribution \citep{2023RAA....23k5011W}. These findings enhance our understanding of the physical mechanisms driving Mrk 421’s variability and contribute to the broader study of blazar jet dynamics.
\subsection{Spectral Analysis}
In the following section, we present the spectral analysis of \textit{NuSTAR} observations of the blazar Mrk 421. This analysis employs a model-independent hardness ratio approach to characterize spectral variability and detailed spectral fitting using various models to describe the \textit{NuSTAR} spectra.

\subsubsection{Hardness Ratio}
To investigate the relationship between the variability in different energy bands of Mrk 421, we extracted light curves in two distinct energy ranges—low-energy (LE, 3–10 keV) and high-energy (HE, 10–79 keV)—following the standard data reduction procedures outlined in Section 2. Our objective was to examine how the high-energy light curve responds to variations in the low-energy light curve, thereby probing the connection between these emission components across the entire observation period. To quantify this relationship, we computed the Spearman’s correlation coefficient between the LE and HE light curves, yielding a value of 0.92 with a p-value of 7.21x10$^{-5}$, which indicates a strong positive correlation between the two energy bands. 
Several studies, utilizing various X-ray telescopes such as XMM-Newton, Swift/XRT, Chandra, and NuSTAR, have reported a similar positive correlation between the emission in the two bands \cite[e. g,][]{2024MNRAS.529.1450G,2022ApJS..262....4N,2018ApJ...854...66K,2018MNRAS.480.4873A,Pandey2017}.

This high degree of correlation suggests that the variability in the LE and HE bands is closely linked, likely driven by common physical processes within the jet. The correlation is visually represented in Figure \ref{fig4}, where we plot the HE light curve against the LE light curve, with each point corresponding to simultaneous flux measurements in the two bands. Additionally, we performed a linear fit to the data, finding that a slope of 0.6 provides the best fit, indicating a proportional but slightly subdued response of the HE emission to changes in the LE emission. For comparison, a dashed red line representing the $x=y$ relation (slope of 1.0) is also included in the figure, highlighting that the HE variability is less sensitive to LE changes. This slope of 0.6 suggests that while the variability in both bands is strongly correlated, the source’s overall variability is dominated by the soft (LE) band, potentially reflecting differences in the electron populations or cooling mechanisms responsible for the emission in these energy ranges \citep{2018MNRAS.481.3563P}.

\begin{figure*}
  \centering
  \includegraphics[width=8cm]{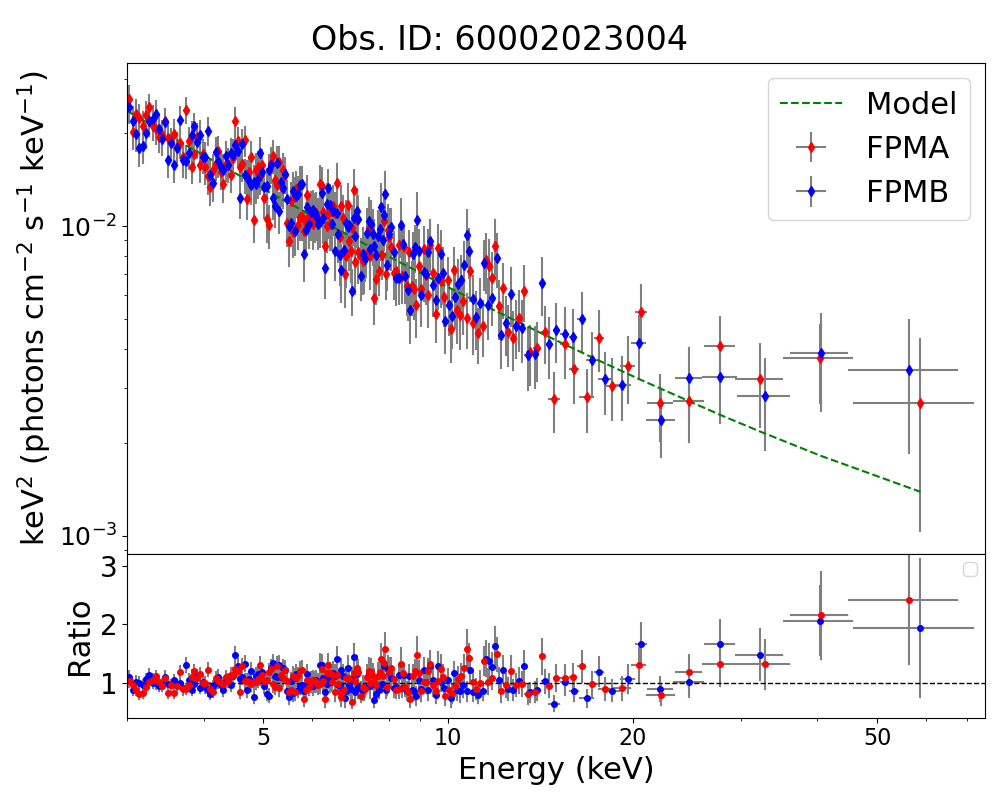}
  \includegraphics[width=8cm]{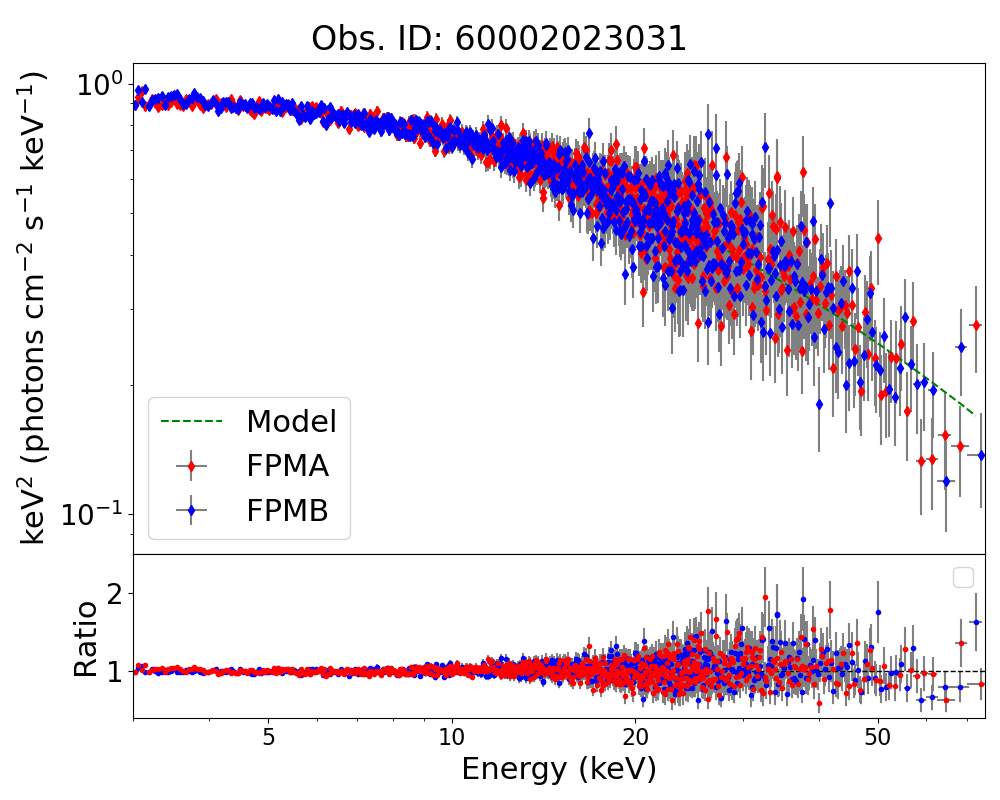}
  \caption{Spectral fitting of \textit{NuSTAR} observations for the lowest (left panel) and highest (right panel) flux states. The spectra were fitted using three XSPEC models: power-law (PL), broken power-law (BPL), and log-parabolic (LP). Among these, the log-parabolic model provides the best fit for both flux states. The ratio plot in the bottom section of the left panel reveals a hard excess during the low flux state at $\sim$ 20~keV.\label{fig5}}
\end{figure*}

To further evaluate the relative contributions of different spectral components to the total flux, as measured by count rates, and to examine how these contributions vary across diverse flux states, we define the hardness ratio (HR) as a model-independent metric of the blazar’s spectral shape. The HR is calculated as:
\begin{equation}\label{eq9}
    \text{HR} = \frac{H}{S},
\end{equation}
where \( H \) represents the flux in the hard X-ray energy band (3–10 keV) and \( S \) denotes the flux in the soft X-ray energy band (10–79 keV). This definition provides a direct measure of the contribution from the hard-band over soft band, facilitating a standardized comparison across observations. The uncertainty in the hardness ratio, \( \sigma_{\text{HR}} \), is derived using error propagation, given by:
\begin{equation}\label{eq6}
    \sigma_{\text{HR}} = \text{HR} \cdot \sqrt{\left( \frac{\sigma_{F_{\text{hard}}}}{F_{\text{hard}}} \right)^2 + \left( \frac{\sigma_{F_{\text{soft}}}}{F_{\text{soft}}} \right)^2},
\end{equation}
where \( \sigma_{F_{\text{hard}}} \) and \( \sigma_{F_{\text{soft}}} \) are the uncertainties in the hard and soft band fluxes, respectively.

The hardness ratio provides a straightforward, model-independent representation of the spectral properties of Mrk 421 within the specified energy range, with variations in HR reflecting spectral changes during the observations. To investigate potential correlations between the spectral properties and the flux state of Mrk 421, we examined the relationship between HR and the total count rate in the entire energy band. The right panel of Figure \ref{fig4} displays the HR values for the full dataset plotted against the logarithm of the count rate, revealing a positive correlation characterized by a slope of 0.05 and a Spearman rank correlation coefficient of 0.48. Harder-when-brighter behavior in Mrk 421 has previously been reported from NuSTAR observations during the 2012–2013 monitoring campaign \citep{2016ApJ...819..156B,Pandey2017}, the 2017 multiwavelength campaign \citep{MAGIC2021,Mondal2022}, and other NuSTAR spectral studies \citep{Middei2022,Baheeja2024}. The present results are broadly consistent with these earlier findings, although the shallow slope indicates that spectral hardening is relatively modest when observations spanning different epochs and activity states are considered together. Such behavior may reflect the combined contribution of emission regions with differing magnetic fields, electron populations, and acceleration processes.

The figure also shows that one observation (ObsID: 60902024010) forms a distinct cluster at a mean count rate of approximately $3\text{--}5\text{ counts s}^{-1}$ and a hardness ratio of $0.33\text{--}0.43$, substantially below the average HR–flux trend. We find that the displacement persists throughout the observation, and we carefully re-examined the analysis to ensure that it reflects source behavior rather than an isolated statistical fluctuation. The observed softness is further supported by the comparatively steep X-ray spectrum independently obtained from the spectral fitting of this observation (see Table \ref{tab2}).
Taken together, these results indicate that Mrk 421 occupied an unusually soft spectral state during this epoch, possibly due to changes in the distribution of emitting electron population and/or magnetic field.


\subsection{Spectral Modeling} 
To model the non-thermal emission from the blazar Mrk 421, we conducted a detailed spectral analysis of 27 \textit{NuSTAR} observations using the \textit{XSPEC} software package, following standard X-ray spectral analysis procedures \citep{Arnaud1996}. During data processing, we generated source spectra, response matrix files (RMF), ancillary response files (ARF), and background spectra for each observation, binning the spectra to ensure a minimum of 20 counts per energy bin for reliable \( \chi^2 \) statistics. The analysis was performed in the 3–79 keV energy range, simultaneously fitting the spectra from the FPMA and FPMB instruments while incorporating an intercalibration constant to account for minor instrumental differences. The intercalibration constants ranged from 0.97 to 1.04, indicating excellent consistency between the two detectors.

We tested three distinct spectral models to determine the best representation of Mrk 421’s spectral behavior: a power law (PL), a log-parabola (LP), and a broken power law (BPL). The power-law model is defined as:
\begin{equation}
\frac{dN}{dE} = N E^{-\Gamma},
\end{equation}
where \( N \), \( E \), and \( \Gamma \) represent the normalization, photon energy, and photon index, respectively. This model reflects the non-thermal emission from a power-law distribution of high-energy particles in the jet’s magnetic field, with the synchrotron spectral index \( \alpha \) related to the particle distribution index \( p \) by \( \alpha = (p - 1)/2 \).

The broken power-law model, which allows for a spectral break, is expressed as:
\begin{equation}
\frac{dN}{dE} = N_0
\begin{cases}
E^{-\Gamma_1}, & \text{if } E \leq E_b \\
E^{-\Gamma_2}, & \text{otherwise}
\end{cases}
\end{equation}
where \( \Gamma_1 \) and \( \Gamma_2 \) are the photon indices below and above the break energy \( E_b \), respectively, and \( N_0 \) is the normalization. A broken power-law spectrum often indicates the presence of high-energy relativistic electrons that cool rapidly through radiative processes, leading to a break in the emission spectrum.

The log-parabola model, characterized by a continuous curvature, is given by:
\begin{equation}
\label{eq10}
\frac{dN}{dE} = N_0 \left( \frac{E}{E_0} \right)^{-\left( \alpha + \beta \log\left( E/E_0 \right) \right)},
\end{equation}
where \( N_0 \) and \( E_0 \) (fixed at 1 keV) are the normalization and reference energy, respectively, while \( \alpha \) and \( \beta \) denote the photon index at \( E_0 \) and the curvature parameter, respectively \citep{Massaro2004a}. The log-parabola model is often associated with particle acceleration scenarios where the probability of further energy gain decreases with increasing particle energy, resulting in a curved spectrum.

To account for Galactic absorption, we incorporated the \emph{tbabs} model (Tuebingen-Boulder ISM absorption model; \citep{Wilms2000}) into each fit, using hydrogen column density values from \cite{Kalberla2005}. For Mrk 421, we fixed the neutral hydrogen column density at \( N_{\text{H}} = 1.34 \times 10^{20} \, \text{cm}^{-2} \), consistent with previous studies \citep{2016A&A...594A.116H}. These spectral models and absorption corrections enable a comprehensive analysis of Mrk 421’s emission properties, providing insights into the physical mechanisms driving its non-thermal radiation in the hard X-ray regime.
The results of the spectral fitting are presented in Table \ref{tab2}. The table includes the observation ID and the XSPEC model in the second and third columns, respectively, while the best fit parameters for the PL, BPL, and LP models are provided in the fourth, fifth, and sixth columns. The final column reports the reduced 
$\chi^2$
  statistic for each fit, enabling a direct comparison of the model performance. The table indicates that the LP model provides the best fit for 20 of the 27 observations of Mrk 421, suggesting that the spectra frequently exhibit curvature. This is consistent with particle acceleration scenarios where the probability of energy gain decreases with increasing energy. The BPL model best fits 2 observations, indicating occasional spectral breaks, possibly due to rapid cooling of high-energy electrons. 
  The spectra of five observations are adequately represented by the PL model. Although the LP and BPL fits yield lower $\chi^2$
 values in these cases, they may overfit the data and are therefore not formally excluded, particularly since the LP model reduces to a PL in the limit $(\beta \to 0)$.
  Nevertheless, the overall preference for the LP and BPL models across the remaining observations indicates that spectral curvature is a common feature of the X-ray emission from Mrk 421 \citep{Massaro2004a}.

Spectral modeling results for two distinct observation epochs of Mrk 421 are presented in Figure \ref{fig5}, with the left panel corresponding to the lowest flux state and the right panel representing the highest flux state. For both observations, the log-parabola (LP) model provided the best fit, as evidenced by the reduced $\chi^2$ values. Notably, the lower panel of the low-state plot reveals a distinct hard excess beginning around 20 keV, indicating a deviation from the LP model at higher energies during this epoch  \citep[see also][]{2016ApJ...827...55K}.

\begin{figure*}[t]
  \centering
  \includegraphics[width=16cm]{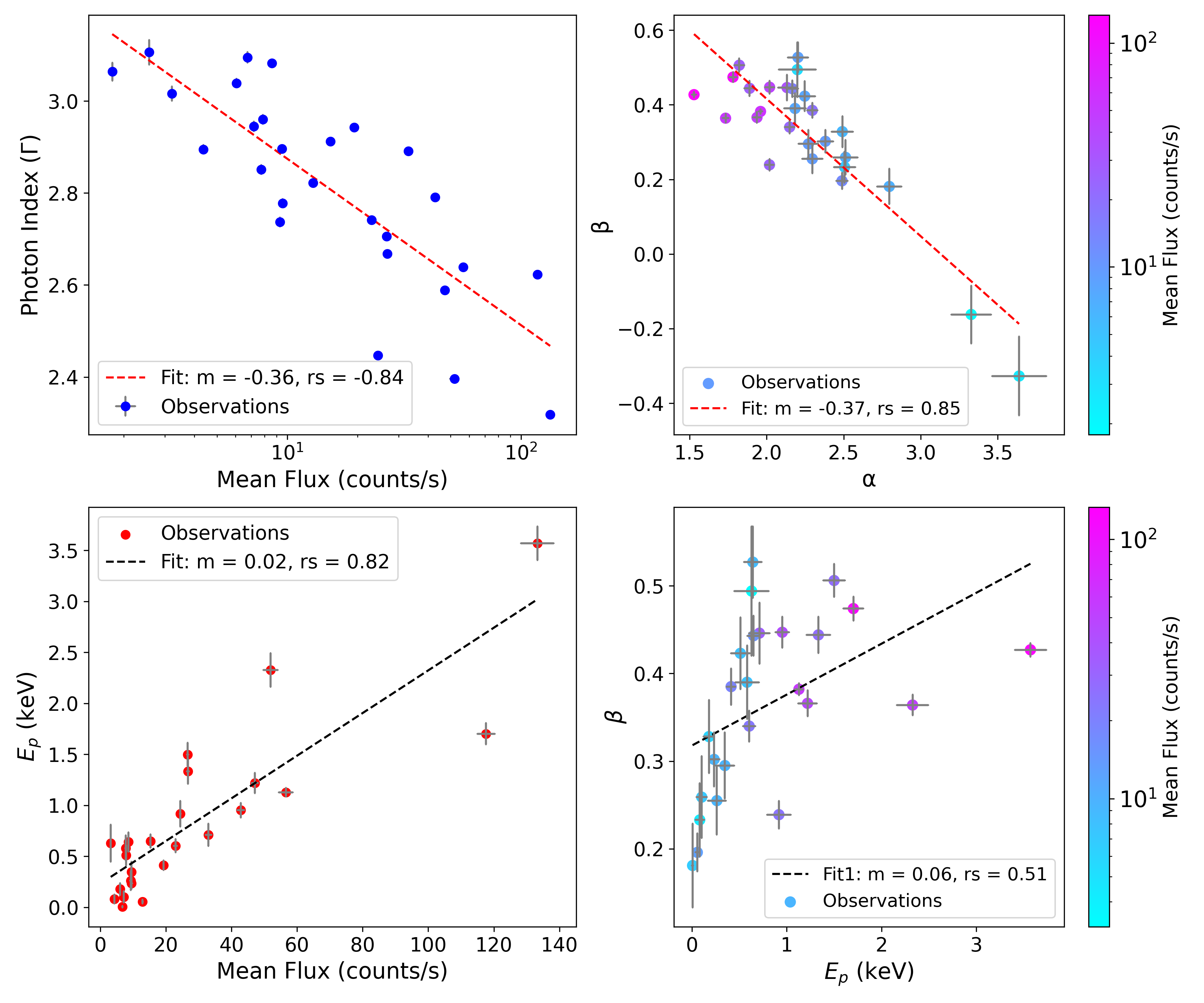}
    \caption{Correlations between spectral parameters from Table~\ref{tab2} and mean count rates from Table~\ref{tab1}, with linear fit parameters and Spearman's correlation coefficients (rs)  provided in the inset text.
  The \textbf{top left} panel shows a strong correlation between the photon index from the power-law (PL) model and the logarithmic mean count rate, indicating a ``harder-when-brighter'' trend.
  The \textbf{top right} panel reveals a strong correlation between the photon index and the curvature parameter from the log-parabolic (LP) model, with negative curvature values suggesting a hard excess in the lowest flux states.
  The \textbf{bottom left} panel displays a strong positive correlation between the peak energy of the LP model and the flux state, indicating that the peak shifts to higher frequencies as the source brightens.
  The \textbf{bottom right} panel shows  a linear fit between the curvature parameter and the LP peak energy. \label{fig6}}
\end{figure*}

To further explore the spectral evolution of Mrk 421 across different flux states, we analyzed the best-fit parameters from Table \ref{tab2}, focusing on their distribution and correlations. The top left panel of Figure \ref{fig6} presents a plot of the photon index from PL model against the logarithm of the mean count rate for each observation (see Table \ref{tab1}), revealing a clear anti-correlation trend that indicates a slow ``harder-when-brighter'' behavior; as the flux increases (on a logarithmic scale), the photon index decreases, suggesting a gradual hardening of the spectrum. The top right panel of the same figure displays an anti-correlation between the \( \alpha \) and \( \beta \) parameters of the LP model, implying that as the spectrum steepens (higher \( \alpha \)), the curvature parameter \( \beta \) decreases, resulting in a flatter spectral shape, with negative \( \beta \) values particularly highlighting a hard excess in the two lowest flux states of the source. Similarly, the bottom left panel illustrates a correlation of mean flux level from each observation with the synchrotron peak energy \( E_p \), derived using the relation \( E_p = E_1 \times 10^{(2-\alpha)/2\beta} \) with \( E_1 = 1 \, \text{keV} \). A positive correlation is observed, with a Spearman's correlation coefficient of 0.82 with a p-value of 2.45x10$^{-7}$. This shift of \( E_p \) toward higher energies as the source brightens further supports the observed spectral hardening during higher flux states.  In a similar previous study, log-parabolic spectral fit to the X-ray observations during the flaring state of the source revealed a  positive correlation between spectral parameters and flux levels \citep{2004ApJ...601..759T, 2009A&A...501..879T}. Furthermore, a correlation study examining the relationship between the peak luminosity and peak energy ($L_{\rm p}$--$E_{\rm p}$ relation) in a sample of 14 BL Lac objects reveals positive, negative, or no correlations between these parameters \cite{2019ApJ...885....8W}. Considering the complete dataset the synchrotron peak energy, 
$E_{\rm p}$, and the curvature parameter, $\beta$, exhibit a moderate 
positive correlation, with $r_{\rm s}=0.51$. The considerable scatter 
indicates that this relation is not uniform across all observational 
epochs and may reflect changes in the particle-acceleration and cooling 
conditions. The black dashed line in Figure~6 represents the fit to all 
observations.

\begin{figure}
\centering
\includegraphics[scale=0.6]{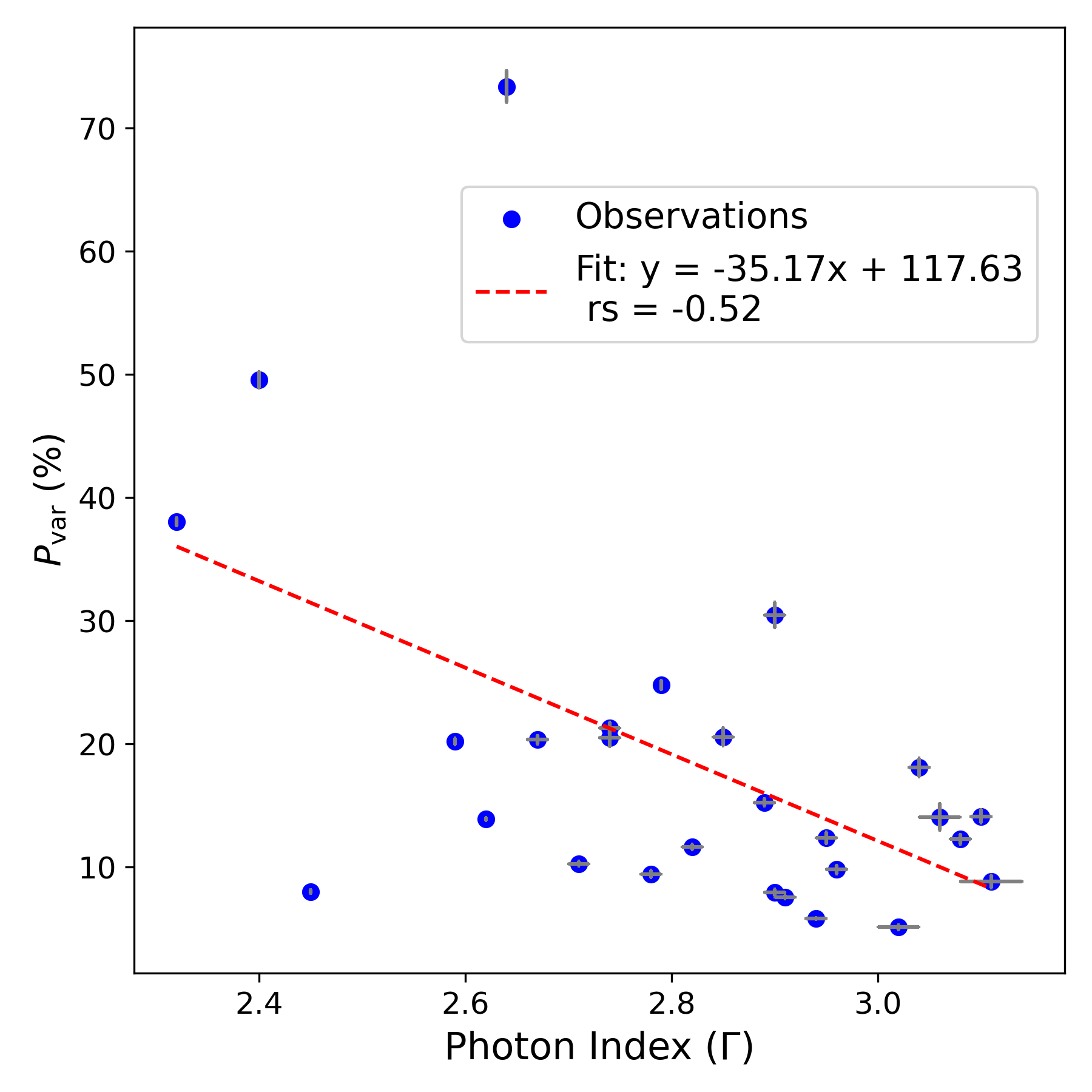}
\caption{ Distribution of variability parameter as a function of photon indices derived from power-law spectral fitting of \textit{NuSTAR} observations of Mrk 421. The red dashed line represents the linear fit to the data. \label{fig7}}
\end{figure}

Similarly, to combine the results of timing and spectral analyses, the variability parameters for each observation are plotted against the photon indices derived from PL spectral fitting in Figure~\ref{fig7}. The plot reveals a trend of anti-correlation between these variables, suggesting greater variability and a broader spread during the harder spectral states of the source.

\section{Discussion}
The study of intra-day and long-term hard X-ray variability in the TeV blazar Mrk 421, based on 12 years of \textit{NuSTAR} observations, reveals flux variations of up to a factor of $\sim 200$. The FV in the X-ray band reaches $\sim$120\%, significantly exceeding the 43\% FV in $\gamma-$rays \cite[][]{{2020ApJ...891..120B}} and 30\% in the optical band \citep[][]{Bhatta2021} over comparable timescales. The source exhibited its highest flux state on April 14, 2013, and its lowest flux state on January 10, 2013. In addition, we observed a profound flaring event in April 2013, during which the source brightened by a factor of 10 within a period of 15 days.  This provides a unique opportunity to probe the physical processes driving blazar variability. By performing comprehensive  spectral and timing analyses—including the linear RMS-flux relation, log-normal flux distributions, and the power-law spectral models—we gain deeper insight into the mechanisms responsible for both intra-day and long-term flux and spectral variability, including this remarkable flaring event. These analyses collectively point to  
\onecolumn
 \begin{strip} 
 {
\setlength{\LTcapwidth}{\textwidth}
\begin{longtable}{ccccccc}
\caption{ Spectral parameters for each observation ID. Serial numbers correspond to those in Table~1. The best-fitting model, characterized by a reduced \(\chi^2 \approx 1.0\), is highlighted in bold fonts.} \label{tab2}\\
\hline
Serial & Observation ID & Model & $\Gamma|\Gamma_1|\alpha$ & $\Gamma_2|\beta$ & $E_b$ & Fit Statistic \\
(1) & (2) & (3) & (4) & (5) & (6) & (7)\\
\hline
\endfirsthead
\caption*{Table \ref{tab2} -- continued} \\
\hline
Serial & Observation ID & Model & $\Gamma|\Gamma_1|\alpha$ & $\Gamma_2|\beta$ & $E_b$ & Fit Statistic \\
\hline
\endhead
\hline
\endfoot
\scriptsize
\setlength{\tabcolsep}{4pt}
 & & PL & $2.896 \pm 0.008$ & ... & ... & 1.21 (806.27/665) \\
1 & 10002016001 & BPL & $1.307 \pm 1.383$ & $2.909 \pm 0.009$ & $3.281 \pm 0.163$ & 1.19 (790.46/663) \\
 & & \textbf{LP} & $2.381 \pm 0.052$ & $0.302 \pm 0.031$ & ... & \textbf{1.05 (699.01/664)} \\
\midrule
 & & PL & $3.106 \pm 0.027$ & ... & ... & 1.16 (303.86/260) \\
2 & 60002023002 & BPL & $0.360 \pm 220.654$ & $3.107 \pm 0.030$ & $2.905 \pm 6.105$ & 1.17 (303.86/258) \\
 & & \textbf{LP} & $3.639 \pm 0.177$ & $-0.327 \pm 0.106$ & ... & \textbf{1.13 (295.19/259)} \\
\midrule
 & & PL & $3.064 \pm 0.020$ & ... & ... & 0.92 (333.29/359) \\
3 & 60002023004 & BPL & $0.448 \pm 53.366$ & $3.067 \pm 0.022$ & $2.983 \pm 1.886$ & 0.92 (331.32/357) \\
 & & \textbf{LP} & $3.328 \pm 0.131$ & $-0.162 \pm 0.078$ & ... & \textbf{0.93 (329.69/358)} \\
\midrule
 & & PL & $3.039 \pm 0.010$ & ... & ... & 1.09 (612.44/558) \\
4 & 60002023006 & BPL & $1.000 \pm 0.000$ & $2.000 \pm 0.000$ & $5.000 \pm 0.000$ & (***) \textsuperscript{a}\\
 & & \textbf{LP} & $2.491 \pm 0.071$ & $0.328 \pm 0.042$ & ... & \textbf{0.98 (547.05/557)} \\
\midrule
 & & PL & $2.960 \pm 0.010$ & ... & ... & 1.16 (663.62/569) \\
5 & 60002023010 & BPL & $2.826 \pm 0.019$ & $3.224 \pm 0.036$ & $8.082 \pm 0.387$ & 0.95 (542.05/567) \\
 & & \textbf{LP} & $2.247 \pm 0.070$ & $0.423 \pm 0.041$ & ... & \textbf{0.95 (543.52/568)} \\
\midrule
 & & \textbf{PL} & $2.737 \pm 0.010$ & ... & ... & \textbf{1.00 (587.58/586)} \\
6 & 60002023012 & BPL & $2.645 \pm 0.019$ & $2.910 \pm 0.035$ & $8.353 \pm 0.630$ & 0.91 (533.84/584) \\
 & & LP & $2.296 \pm 0.067$ & $0.255 \pm 0.039$ & ... & 0.92 (541.22/585) \\
\midrule
 & & \textbf{PL} & $3.095 \pm 0.012$ & ... & ... & \textbf{0.91 (463.67/506)} \\
7 & 60002023018 & BPL & $0.538 \pm 5.973$ & $3.105 \pm 0.013$ & $3.135 \pm 0.316$ & 0.90 (458.56/504) \\
 & & LP & $2.796 \pm 0.080$ & $0.181 \pm 0.048$ & ... & 0.88 (448.56/505) \\
\midrule
 & & \textbf{PL} & $2.778 \pm 0.009$ & ... & ... & \textbf{1.01 (600.08/591)} \\
8 & 60002023020 & BPL & $2.693 \pm 0.017$ & $2.988 \pm 0.037$ & $9.034 \pm 0.605$ & 0.90 (534.76/589) \\
 & & LP & $2.271 \pm 0.065$ & $0.295 \pm 0.038$ & ... & 0.90 (531.65/590) \\
\midrule
 & & PL & $2.741 \pm 0.005$ & ... & ... & 1.47 (1320.83/895) \\
9 & 60002023022 & BPL & $2.650 \pm 0.008$ & $3.008 \pm 0.021$ & $9.842 \pm 0.298$ & 1.07 (958.93/893) \\
 & & \textbf{LP} & $2.149 \pm 0.032$ & $0.340 \pm 0.018$ & ... & \textbf{1.04 (930.27/894)} \\
\midrule
 & & PL & $2.891 \pm 0.008$ & ... & ... & 1.34 (835.71/621) \\
10 & 60002023024 & BPL & $2.772 \pm 0.014$ & $3.198 \pm 0.035$ & $8.956 \pm 0.375$ & 1.07 (667.26/619) \\
 & & \textbf{LP} & $2.132 \pm 0.060$ & $0.446 \pm 0.035$ & ... & \textbf{1.05 (652.12/620)} \\
\midrule
 & & PL & $2.639 \pm 0.002$ & ... & ... & 3.30 (4713.86/1426) \\
11 & 60002023025 & BPL & $2.524 \pm 0.003$ & $2.928 \pm 0.008$ & $9.793 \pm 0.106$ & 1.18 (1681.61/1424) \\
 & & \textbf{LP} & $1.960 \pm 0.013$ & $0.382 \pm 0.007$ & ... & \textbf{1.01 (1451.51/1425)} \\
\midrule
 & & PL & $2.623 \pm 0.004$ & ... & ... & 2.29 (2352.14/1025) \\
12 & 60002023027 & BPL & $2.484 \pm 0.006$ & $2.994 \pm 0.017$ & $10.029 \pm 0.169$ & 1.06 (1088.72/1023) \\
 & & \textbf{LP} & $1.781 \pm 0.025$ & $0.474 \pm 0.014$ & ... & \textbf{0.98 (1012.73/1024)} \\
\midrule
 & & PL & $2.791 \pm 0.004$ & ... & ... & 1.75 (1612.25/917) \\
13 & 60002023029 & \textbf{BPL} & $2.629 \pm 0.009$ & $3.039 \pm 0.014$ & $7.855 \pm 0.160$ & \textbf{0.97 (891.22/915)} \\
 & & LP & $2.019 \pm 0.030$ & $0.447 \pm 0.018$ & ... & 0.92 (850.97/916) \\
\midrule
 & & PL & $2.318 \pm 0.002$ & ... & ... & 3.42 (4931.28/1440) \\
14 & 60002023031 & BPL & $2.183 \pm 0.004$ & $2.690 \pm 0.010$ & $11.264 \pm 0.127$ & 1.14 (1648.47/1438) \\
 & & \textbf{LP} & $1.528 \pm 0.015$ & $0.427 \pm 0.008$ & ... & \textbf{1.06 (1531.32/1439)} \\
\midrule
 & & PL & $2.589 \pm 0.004$ & ... & ... & 1.71 (1750.38/1020) \\
15 & 60002023033 & BPL & $2.475 \pm 0.007$ & $2.867 \pm 0.016$ & $9.747 \pm 0.216$ & 1.02 (1038.80/1018) \\
 & & \textbf{LP} & $1.937 \pm 0.026$ & $0.366 \pm 0.015$ & ... & \textbf{0.99 (1014.38/1019)} \\
\midrule
 & & PL & $2.396 \pm 0.003$ & ... & ... & 1.93 (2309.15/1191) \\
16 & 60002023035 & BPL & $2.283 \pm 0.005$ & $2.685 \pm 0.014$ & $10.517 \pm 0.200$ & 1.04 (1242.80/1189) \\
 & & \textbf{LP} & $1.733 \pm 0.021$ & $0.364 \pm 0.012$ & ... & \textbf{0.98 (1168.20/1190)} \\
\midrule
 & & PL & $2.851 \pm 0.010$ & ... & ... & 1.09 (619.21/566) \\
17 & 60002023037 & \textbf{BPL} & $2.740 \pm 0.018$ & $3.114 \pm 0.040$ & $8.754 \pm 0.489$ & \textbf{0.93 (524.59/564)} \\
 & & LP & $2.184 \pm 0.071$ & $0.390 \pm 0.042$ & ... & 0.91 (516.76/565) \\
\midrule
 & & \textbf{PL} & $2.945 \pm 0.011$ & ... & ... & \textbf{1.02 (535.31/524)} \\
18 & 60002023039 & BPL & $2.764 \pm 0.043$ & $3.045 \pm 0.024$ & $5.577 \pm 0.414$ & 0.94 (494.59/522) \\
 & & LP & $2.512 \pm 0.080$ & $0.259 \pm 0.047$ & ... & 0.96 (503.61/523) \\
\midrule
 & & PL & $2.447 \pm 0.004$ & ... & ... & 1.29 (1298.79/1006) \\
19 & 60202048002 & BPL & $2.353 \pm 0.009$ & $2.589 \pm 0.013$ & $8.450 \pm 0.319$ & 1.06 (1067.30/1004) \\
 & & \textbf{LP} & $2.018 \pm 0.029$ & $0.239 \pm 0.016$ & ... & \textbf{1.04 (1054.17/1005)} \\
\midrule
 & & PL & $3.082 \pm 0.009$ & ... & ... & 1.40 (800.31/570) \\
20 & 60702061002 & BPL & $2.904 \pm 0.020$ & $3.354 \pm 0.030$ & $7.313 \pm 0.275$ & 1.07 (610.69/568) \\
 & & \textbf{LP} & $2.203 \pm 0.067$ & $0.527 \pm 0.041$ & ... & \textbf{1.05 (597.93/569)} \\
\midrule
 & & PL & $2.706 \pm 0.005$ & ... & ... & 2.08 (1845.52/887) \\
21 & 60702061004 & BPL & $2.550 \pm 0.008$ & $3.048 \pm 0.019$ & $9.022 \pm 0.179$ & 1.16 (1030.93/885) \\
 & & \textbf{LP} & $1.822 \pm 0.033$ & $0.506 \pm 0.019$ & ... & \textbf{1.11 (984.94/886)} \\
\midrule
 & & \textbf{PL} & $2.668 \pm 0.005$ & ... & ... & \textbf{0.63 (811.57/1282)} \\
22 & 80801643002 & BPL & $2.509 \pm 0.011$ & $2.923 \pm 0.017$ & $8.229 \pm 0.207$ & 0.24 (312.75/1280) \\
 & & LP & $1.889 \pm 0.036$ & $0.444 \pm 0.021$ & ... & 0.20 (265.83/1281) \\
\midrule
 & & PL & $2.822 \pm 0.006$ & ... & ... & 1.07 (880.85/820) \\
23 & 60902024002 & BPL & $0.523 \pm 5.007$ & $2.828 \pm 0.006$ & $3.092 \pm 0.265$ & 1.06 (873.67/818) \\
 & & \textbf{LP} & $2.489 \pm 0.038$ & $0.196 \pm 0.022$ & ... & \textbf{0.97 (799.67/819)} \\
\midrule
 & & PL & $2.912 \pm 0.005$ & ... & ... & 1.56 (1254.12/799) \\
24 & 60902024004 & BPL & $2.735 \pm 0.014$ & $3.107 \pm 0.015$ & $6.781 \pm 0.176$ & 1.08 (865.23/797) \\
 & & \textbf{LP} & $2.166 \pm 0.039$ & $0.443 \pm 0.023$ & ... & \textbf{1.03 (829.40/798)} \\
\midrule
 & & PL & $2.943 \pm 0.005$ & ... & ... & 1.43 (1219.41/847) \\
25 & 60902024006 & BPL & $2.810 \pm 0.011$ & $3.135 \pm 0.015$ & $7.304 \pm 0.203$ & 1.02 (869.68/845) \\
 & & \textbf{LP} & $2.296 \pm 0.035$ & $0.385 \pm 0.021$ & ... & \textbf{0.98 (829.82/846)} \\
\midrule
 & & PL & $3.016 \pm 0.016$ & ... & ... & 4.92 (2554.74/519) \\
26 & 60902024010 & BPL & $2.848 \pm 0.034$ & $3.279 \pm 0.054$ & $7.248 \pm 0.499$ & 4.83 (2501.70/517) \\
 & & \textbf{LP} & $2.199 \pm 0.122$ & $0.494 \pm 0.074$ & ... & \textbf{4.83 (2504.25/518)} \\
\midrule
 & & PL & $2.895 \pm 0.010$ & ... & ... & 1.22 (700.40/574) \\
27 & 60902024012 & BPL & $0.533 \pm 6.650$ & $2.901 \pm 0.011$ & $3.116 \pm 0.364$ & 1.22 (698.96/572) \\
 & & \textbf{LP} & $2.506 \pm 0.071$ & $0.233 \pm 0.042$ & ... & \textbf{1.16 (669.90/573)} \\
\hline
\multicolumn{7}{l}{%
\footnotesize
\textsuperscript{a} The model failed to converge during the fitting process.
} \\
\end{longtable}
}

\twocolumn
\end{strip}
\noindent multiplicative, non-linear processes within the relativistic jet, likely driven by enhanced particle injection at a shock wavefront, with possible contributions from magnetic reconnection or accretion disk fluctuations. 

The hard X-ray emission during the observation period exhibits distinct spectral characteristics. The nearly constant hardness ratio $\mathrm{HR}=F_{\rm HE}/F_{\rm LE}$, with a slope of 0.05 in the $\log F$ versus HR relation) and the strong linear correlation between the high-energy (HE; 10–79 keV) and low-energy (LE; 3–10 keV) fluxes $F_{\rm HE}\approx0.6F_{\rm LE}$ indicate that the variations in the two bands are closely coupled and accompanied by little spectral evolution. Within a leptonic interpretation, this behaviour is consistent with both bands tracing synchrotron emission from a common electron population with an approximately stable energy distribution, although it does not by itself exclude alternative emission scenarios. Such spectral stability could arise if particle acceleration and radiative losses approach a quasi-steady equilibrium. For synchrotron-dominated losses, the cooling timescale scales as $t_{\rm syn}\propto(B^{2}\gamma_{\rm e})^{-1}$, where (B) is the magnetic-field strength in the emitting region and $\gamma_{\rm e}$ is the Lorentz factor of the radiating electrons \citep{Kirk1998,Stawarz2008,Tramacere2011}. Therefore acceleration–cooling balance can be regarded as a probable explanation for the observed spectral stability, although other physical mechanisms cannot be ruled out.

The log-parabolic spectral model, expressed by Equation \ref{eq10}, reveals a significant anti-correlation between the photon index \( \alpha \) (2.38--3.64) and curvature parameter \( \beta \) (-0.33--0.30), in contrast  the positive correlation observed in other X-ray observations of the source \citep[e. g.,][]{Massaro2004a, 2018ApJ...854...66K}. This discrepancy 

\noindent suggests that
  during the period of hard X-ray emission, particle acceleration is influenced by additional processes, such as stochastic acceleration or energy-dependent cooling, which alter the electron energy distribution and spectral shape. The synchrotron peak energy, calculated as \( E_p = E_1 \times 10^{(2-\alpha)/(2\beta)} \) with \( E_1 = 1 \, \text{keV} \) \citep[see][]{2004A&A...422..103M}, shifts from \( \approx 0.23 \, \text{keV} \) (\( \alpha = 2.38 \), \( \beta = 0.30 \)) to \( \approx 3.57 \, \text{keV} \) (\( \alpha = 2.51 \), \( \beta = 0.53 \)) as the flux increases, aligning with a "harder-when-brighter" trend (photon index \( \Gamma \) decreases from 3.1 to 2.3 as mean count rate rises from 1.76 to 133.25 counts/s).

 This spectral evolution can be interpreted in terms of changes in the characteristic electron Lorentz factor, $\gamma_{\rm p}$, the magnetic-field strength, $B$, and the normalization of the radiating electron population, $K_{\rm e}$. For synchrotron emission, the observed peak frequency scales as $\nu_{\rm p}\propto B\gamma_{\rm p}^{2}\delta_{\rm D}/(1+z)\ {\rm Hz}$, where $z$ and $\delta_{\rm D}$ represent the red-shift and the Doppler factor, respectively \citep{Tavecchio1998}. The radiated power of an individual electron is  $P_{\rm syn}\propto B^2\gamma^2$; consequently, for an optically thin synchrotron spectrum approximated locally by a power law, the observed flux scales as $F_{\nu}\propto K_{\rm e}B^{1+\alpha_{\rm X}}\delta_{\rm D}^{3+\alpha_{\rm X}}$, with $\alpha_{\rm X}=\Gamma-1$, where $\Gamma$ is the photon index. These relations show that particle acceleration accompanied by a modest enhancement of the magnetic field can increase both the synchrotron peak energy and the observed X-ray flux. Such magnetic-field amplification can result from shock compression \citep{Zhang2014}, while magnetic reconnection can accelerate particles and produce localized magnetized emission regions \citep{Petropoulou2016,Christie2019}. Therefore, a magnetically influenced response provides a physically plausible interpretation of the observed harder-when-brighter behavior, although contributions from variations in other parameters, e.g., $B$, $K_{\rm e}$, and $\gamma_{\rm p}$ can not be ruled out. This interpretation is further supported by the relatively stable HR and the approximately linear correlation between the low- and high-energy count rates, suggesting that both bands are dominated by the same synchrotron component.
In a simple one-zone SSC scenario in the Thomson regime, variations driven primarily by the electron density are expected to produce a nonlinear relation, $F_{\rm HE}\propto F_{\rm LE}^{2}$ \citep{2005A&A...433..479K}, while a substantial hadronic contribution may produce stronger spectral hardening with increasing flux \citep{2019MNRAS.483L..12C,2015MNRAS.447...36P}.

Changes in the Doppler factor, $\delta_{\rm D}=[\Gamma(1-\beta\cos\theta)]^{-1}$, where $\Gamma$ is the bulk Lorentz factor, $\beta c$ is the jet velocity, and $\theta$ is the viewing angle, can also influence the observed flux and synchrotron peak energy. For a discrete emitting region with an intrinsic power-law spectrum, $F'_{\nu'}\propto\nu'^{-\alpha_{\rm X}}$, relativistic beaming gives $F_{\nu}\propto\delta_{\rm D}^{3+\alpha_{\rm X}}$, while the observed synchrotron peak energy scales as $E_{\rm p}\propto\delta_{\rm D}$ when the intrinsic emission properties remain unchanged \citep{LindBlandford1985,Urry1995}. For the observed range $\alpha_{\rm X}=\Gamma-1\simeq1.3\text{--}2.1$, a tenfold increase in flux would therefore require $\delta_{\rm D}$ to increase by only a factor of $10^{1/(3+\alpha_{\rm X})}\simeq1.6\text{--}1.7$. Such a change would shift $E_{\rm p}$ by the same factor, substantially less than the reported order-of-magnitude displacement. Therefore, Doppler-factor variations alone are unlikely to explain the simultaneous flux and spectral evolution, suggesting that changes in the electron population and/or magnetic field also contribute.

The timing properties of Mrk 421’s flare further shed light on the underlying variability mechanisms. The linear trend in RMS-flux relation, where the root-mean-squared (RMS) amplitude of variability generally scales linearly with the mean flux, indicates that higher flux states exhibit greater variability, suggesting a common mechanism linking short- and long-timescale variations. For Mrk 421,  this relation in X-ray light curves from \textit{AstroSat} and \textit{XMM-Newton} across epochs spanning decades and shorter periods (5 days to \( \sim \)10 years), similar to patterns in X-ray binaries and Seyfert galaxies \cite{Bhattacharyya2020,2004ApJ...612L..21G,2005MNRAS.359..345U}. A broader study involving a large population of blazars, using 14 years of \textit{Fermi}-LAT gamma-ray light curves, found that a significant number of sources—particularly BL Lac objects like Mrk 421—exhibited a strong linear RMS-flux relation. This relation was interpreted in terms of internal shock models with variations in shock amplitudes \citep{ 2023RAA....23k5011W}. However, the generality of the relation is debated, as some simulations suggest it appears in only 26--32\% of light curves under specific shock scenarios \citep{2022MNRAS.510.3688K}.

Complementing the RMS–flux relation, the analysis of \textit{NuSTAR} light curves reveals a log-normal probability density function (PDF), with a reduced 
\( \chi^2 \sim 7 \), significantly better than that of a normal distribution (\( \chi^2 \sim 35 \)). In the AGN literature, these characteristics are considered indicative of multiplicative, non-linear processes. This log-normality, also reported in \textit{AstroSat} and \textit{XMM-Newton} X-ray data for Mrk 421 \citep{Bhattacharyya2020} and in \textit{Fermi}-LAT gamma-ray data for 46.5\% of BL Lacs \citep{2023RAA....23k5011W}, suggests turbulent or cascading energy injection in the jet \citep{Shah2018}. In contrast, flat-spectrum radio quasars (FSRQs) like 3C 273 occasionally show normal distributions, highlighting diverse variability mechanisms \citep{Bhattacharyya2020}.
The timing analyses, together with spectral analyses, present a coherent picture of the flare in Mrk 421. The linear RMS-flux relation and log-normal flux distribution point to multiplicative, non-linear processes, likely driven by internal shocks or turbulent energy injection within the jet. These processes align with the spectral findings, where the anti-correlation between \( \alpha \) and \( \beta \), coupled with the "harder-when-brighter" trend and \( E_p \) shift, suggests complex particle acceleration mechanisms, possibly involving stochastic acceleration or energy-dependent cooling. The stable HR and linear flux correlation underscore the dominance of leptonic synchrotron radiation, with internal shocks increasing \( \gamma_{e,\text{max}} \) and \( N_e \), potentially amplified by a modest increase in \( B \).

This variability may originate from accretion rate fluctuations in the disk, which propagate along the jet and carry imprints of disk properties, constraining the disk-jet connection \citep{2009A&A...503..797G, 2020ApJ...891..120B, Bhatta2021}. The anti-correlation between \( \alpha \) and \( \beta \), deviating from standard models, further suggests that additional factors, such as magnetic reconnection or spatially varying acceleration regions, modulate the electron energy distribution during the flare. While deviations from log-normality in FSRQs and the non-universal RMS-flux relation indicate complexity, the consistent log-normal PDFs and RMS-flux relation in Mrk 421 reinforce a turbulent jet origin for its variability.

The spectral and timing analyses of the Mrk 421 flare reveal a dynamic interplay of multiplicative, non-linear processes driving its variability. The log-parabolic spectral model, with an anti-correlation between \( \alpha \) and \( \beta \), points to complex particle acceleration mechanisms, while the "harder-when-brighter" trend and \( E_p \) shift suggest enhanced particle injection at a shock wavefront, possibly coupled with magnetic reconnection. The linear RMS-flux relation and log-normal flux distribution support these findings, indicating turbulent energy injection and potential disk-jet coupling. These results highlight the dominance of leptonic synchrotron processes in Mrk 421’s jet and underscore the need for further multiwavelength studies to refine our understanding of blazar jet dynamics and their connections to the central engine.

The spectral and timing analyses of the hard X-ray variability in the TeV blazar Mrk 421, based on 12 years of \textit{NuSTAR} observations, provide a coherent picture of the underlying physical mechanisms driving its variability. The significant flux variations (up to a factor of 225, with a variability parameter of $\sim$127\% in X-rays) and the "harder-when-brighter" trend, evidenced by the observed shift in synchrotron peak energy, suggest enhanced particle injection at a shock wavefront, likely amplified by a modest increase in the magnetic field (\textit{B}). The log-parabolic spectral model's anti-correlation between the photon index $\alpha$ (2.38--3.64) and curvature parameter $\beta$ (-0.33--0.30), contrasting with prior observations \citep{Massaro2004a, 2018ApJ...854...66K}, indicates complex particle acceleration processes, such as stochastic acceleration or energy-dependent cooling. The linear RMS-flux relation and log-normal flux distribution, with a reduced $\chi^2 \sim 7$ compared to $\chi^2 \sim 35$ for a normal distribution, point to multiplicative, non-linear processes, consistent with turbulent energy injection or internal shocks within the relativistic jet \citep{Shah2018, 2023RAA....23k5011W}. The stable hardness ratio and linear correlation between high- and low-energy fluxes further support the dominance of leptonic synchrotron radiation, ruling out significant synchrotron self-Compton or hadronic contributions. These findings, potentially linked to accretion disk fluctuations propagating through the jet \citep{2009A&A...503..797G}, underscore a dynamic interplay of processes in Mrk 421's jet.

\section{Summary and conclusions}
We conducted a comprehensive timing and spectral analysis of the TeV blazar Mrk 421, utilizing 27 observations spanning 12 years from 2012 to 2024, to investigate its hard X-ray variability properties. Our key findings are summarized as follows:

\begin{enumerate}
\item Mrk 421 exhibited extreme variability, with flux changes of up to a factor of 200 over a 12-year period, including a 15-day flaring event during which the source brightened by more than a factor of 10. This dramatic variability is likely driven by an increase in the maximum Lorentz factor during shock acceleration, highlighting the dynamic nature of the particle acceleration processes in the jet.
\item The hard X-ray variability displays a weak  linear trend of RMS-flux relation and a lognormal flux distribution. These characteristics suggest that the variability arises from multiplicative processes, rather than additive ones, pointing to a cascading or interconnected mechanism governing the emission.
\item A strong correlation is observed between the count rates in the low-energy (3--10 keV) and high-energy (10--79 keV) bands, indicating that synchrotron emission dominates the X-ray output across these energy ranges.
\item Spectral analysis reveals a weak but consistent harder-when-brighter trend, with spectra best described by log-parabolic models exhibiting curvature and occasional breaks. The photon index and curvature parameters of the log-parabola model are anti-correlated, and the spectral peak shifts to higher energies with increasing flux. These features collectively support a scenario of energy-dependent stochastic particle acceleration shaping the emission.
\end{enumerate}

These results provide valuable insight into the physical processes driving Mrk 421’s variability, underscoring the interplay of particle acceleration and synchrotron emission in blazar jets. Our findings pave the way for further studies to refine models of relativistic jet dynamics and their observational signatures.

\section*{Acknowledgments}
We are grateful to the anonymous reviewer for their thoughtful and constructive comments, which helped significantly improve the manuscript.
We sincerely thank Ms. Sangeetha Kizhakkekalam for her careful reading of the manuscript and for the valuable discussions.
This work was partially supported by
a program of the Polish Ministry of Science under the title ``Regional Excellence Initiative”, project No. RID/SP/0050/2024/1.

\section*{Data Availability}

The NuSTAR data used in this paper are publicly available and can be accessed through the High Energy Astrophysics Science Archive Research Center (HEASARC) at \url{https://heasarc.gsfc.nasa.gov}.




\bibliographystyle{mnras}
\bibliography{example} 





\bsp	
\label{lastpage}
\end{document}